\documentclass[12pt]{article}
\usepackage{amsmath, amssymb, slashed, epsf, color, graphicx}
\usepackage{epsfig}

\def\dslash{\partial \hspace*{-0.5em}/\hspace*{0.3em}}

\begin{document}

%%%%%%%%%%%%%%%%%%%%%%
%%%%% Title page %%%%%
%%%%%%%%%%%%%%%%%%%%%%
\begin{titlepage}
\begin{center}

\hfill IPMU13-0127 \\
\hfill KEK-TH-1639 \\

\vspace{1.5cm}
{\large\bf Phenomenology of Light Fermionic Asymmetric Dark Matter}

\vspace{2.0cm}
{\bf Biplob Bhattacherjee}$^{(a)}$,
{\bf Shigeki Matsumoto}$^{(a)}$, \\
{\bf Satyanarayan Mukhopadhyay}$^{(a)}$
{\bf and}
{\bf Mihoko M. Nojiri}$^{(a, b)}$ \\

\vspace{1.0cm}
{\it
$^{(a)}${Kavli IPMU (WPI), University of Tokyo, Kashiwa, Chiba 277-8583, Japan} \\
$^{(b)}${KEK Theory Center and Sokendai, Tsukuba, Ibaraki 305-0801, Japan}} \\[2cm]

\vspace{1.0cm}
\abstract{
Asymmetric dark matter (ADM) has been an attractive possibility attempting to explain the observed ratio of baryon to dark matter abundance in the universe. While a bosonic ADM is constrained by the limits from existence of old neutron stars, a fermionic ADM requires an additional light particle in order to annihilate its symmetric component in the early universe. We revisit the phenomenology of a minimal GeV scale fermionic ADM model including a light scalar state. The current constraints on this scenario from cosmology, dark matter direct detection, flavour physics and collider searches are investigated in detail. We estimate the future reach on the model parameter space from next-generation dark matter direct detection experiments, Higgs boson property measurements and search for light scalars at the LHC, as well as the determination of Higgs invisible branching ratio at the proposed ILC.}

\end{center}
\end{titlepage}
\setcounter{footnote}{0}

%%%%%%%%%%%%%%%%%%%%%%%%
%%%%% Introduction %%%%%
%%%%%%%%%%%%%%%%%%%%%%%%
\section{Introduction}
\label{sec: introduction}

The discovery of a scalar boson at the Large Hadron Collider (LHC) experiment~\cite{ATLAS,CMS}, whose properties are broadly consistent with a standard model (SM) Higgs boson, has provided us with the last missing piece in the SM. On the other hand, so far, there is no clear signal of new physics beyond the SM (BSM) either at the LHC or at low energy flavour sector experiments. However, we still need BSM physics to address the long-standing questions on neutrino masses and mixings, the baryon asymmetry of the universe (BAU), and the existence of dark matter (DM)\footnote{In the cosmic frontier, the inflationary paradigm, which is strongly favoured by the WMAP~\cite{WMAP} and recent Planck~\cite{Planck} results, also requires the existence of BSM physics.}. It is a well-known fact that the existence of heavy Majorana neutrinos can simultaneously explain the smallness of neutrino masses via the see-saw mechanism~\cite{see-saw}, and the BAU of the universe via leptogenesis~\cite{lepto}. Right-handed neutrinos are also necessary for cancellation of anomalies, if there exists a gauged U(1)$_{\rm B-L}$ symmetry, which is expected to be spontaneously broken down at a high scale.

In various extensions of the SM, dark matter is accommodated as a stable thermal relic\footnote{Primordial black-holes, produced during the inflationary epoch, remain a viable candidate for dark matter, and do not require the introduction of new particles.}. One of the most important questions in this regard is which symmetry guarantees the stability of the DM particle. An attractive answer to this question is a residual symmetry of the gauged U(1)$_{\rm B-L}$~\cite{IMY}. Since the right-handed neutrinos acquire their Majorana mass terms when this symmetry is broken by the vacuum expectation value of a field with a B$-$L charge of two, a $Z_2$ symmetry remains as a residual one. Thus, a new fermion with an even B$-$L charge or a new boson with an odd B$-$L charge becomes automatically stable, since the physical states in the SM are either fermions with an odd B$-$L charge or bosons with an even B$-$L charge. A new particle with a fractional B$-$L charge will also be stable.

We now briefly recall the thermal history of a stable particle carrying a B$-$L charge, which can act as a DM candidate (for details on the thermal history we refer the reader to reference~\cite{IMY}). In the early universe, the dark matter particle is expected to be in chemical and thermal equilibrium with the SM sector. Since the SM particles should develop a B$-$L asymmetry consistent with the BAU observed today (B$-$L asymmetry is assumed to be generated via leptogenesis in the very early universe), there must also be an asymmetry between dark and anti-dark matter particles. When the temperature of the universe becomes low enough, the annihilation between dark and anti-dark matter particles starts eliminating their symmetric component, and eventually either of the dark or anti-dark matter particles survives until today. To realize such a scenario, the annihilation cross section between dark and anti-dark matter particles in the early universe must be large enough. The dark matter particle carrying a B$-$L charge may therefore be regarded as an asymmetric dark matter (ADM), which has been frequently discussed in past studies~\cite{ADM}.

If we assume that the asymmetry transfer decouples at the time when ADM is relativistic,  the ADM mass can be definitely predicted by its B$-$L charge without depending on the details of its interactions~\cite{IMY}, and is given by $m_{\rm DM} \simeq 5.7 \, {\rm GeV}/Q_{\rm DM}$, where $Q_{\rm DM}$ is the B$-$L charge of the ADM particle. The ADM must be a singlet under the SM gauge group when $Q_{\rm DM} = {\cal O}(1)$, since otherwise it is in conflict with the invisible decay width of the $Z$ boson measured at the Large Electron-Positron Collider (LEP) experiment. There is another severe constraint on the ADM scenario when it is a scalar boson. As the self-annihilation cross section between dark matter particles or between anti-dark matter particles is highly suppressed at the present universe, ADM particles keep being accumulated inside neutron stars, eventually forming black holes inside the stars and destroying them~\cite{Scalar ADM}. As a result, the observation of old neutron stars gives a very severe limit on the scattering cross section between the scalar ADM and a nucleon. It is not easy to consider a mechanism to suppress such interactions, since the dark matter always has a renormalizable interaction $|\phi|^2 |H|^2$ with $\phi$ and $H$ being the ADM and Higgs fields, respectively.

In this paper, we therefore focus on a light fermionic ADM scenario. Since the ADM must be a singlet under the SM gauge group and thus does not have any renormalizable interaction with the SM particles, it requires an additional light particle (called a light mediator) in order to eliminate its symmetric component in the early universe. The introduction of a new light scalar particle, which is also a singlet under the SM gauge group, gives the minimal setup for the fermionic ADM scenario, where the scalar particle does not introduce any dangerous flavor changing processes. Interestingly, the existence of such a light scalar particle does affect various phenomena at both high-energy (collider) and low-energy experiments. Furthermore, since the scalar particle plays a role to connect the ADM to the SM sector, dark matter physics is also affected by the scalar\footnote{If the mediator mass is much smaller than the DM mass, there can be constraints coming from the observation of old neutron stars~\cite{Kouvaris}. However, such a scenario is also excluded by DM halo shape constraints, as shown later in figure~\ref{fig: P}.}. It is therefore important to perform a comprehensive analysis of the fermionic ADM scenario with a light scalar mediator. 

After briefly reviewing the light ADM scenario and describing our setup in the next section, we discuss cosmological, collider and low-energy constraints in section~\ref{sec: constraints} and clarify the parameter region of the model which is allowed by these constraints. In section~\ref{sec: prospects}, we investigate how this parameter region can be explored at on-going and (near) future experiments. We see that direct detection experiments for dark matter and invisible Higgs decay searches at collider experiments such as the LHC and the International Linear Collider (ILC) will play the most important role for this purpose. Section~\ref{sec: summary} is devoted to a summary of our study.

%%%%%%%%%%%%%%%%%
%%%%% Setup %%%%%
%%%%%%%%%%%%%%%%%
\section{The minimal fermionic ADM model}
\label{sec: setup}

\subsection{The ADM scenario}

Before describing the minimal ADM model used in our analysis, we briefly summarize the ADM scenario with a focus on the relation between the mass and the B$-$L charge of the ADM particle. As already mentioned in the introduction, the ADM particle is assumed to be in thermal and chemical equilibrium with the SM sector in the early universe. The interaction maintaining the equilibrium can be described by an effective interaction of the following form\footnote{The equilibrium between the DM and the SM sectors can be achieved also via new sphaleron processes associated with an extra non-abelian gauge symmetry common to both the sectors~\cite{Blennow}.}:
\begin{eqnarray}
{\cal L}_{\rm int}
=
\frac{1}{\Lambda^n}{\cal O}_{\rm DM} \cdot {\cal O}_{\rm SM} + {\rm h.c.},
\label{eq: ADM interaction}
\end{eqnarray}
where ${\cal O}_{\rm DM}$ involves only dark and anti-dark matter fields and needs to be DM number violating, while ${\cal O}_{\rm SM}$ consists of SM fields. This interaction should preserve the total B$-$L number, since otherwise the B$-$L asymmetry generated in the very early universe will be washed out. The baryon asymmetry of the universe is, as a result, related to the B$-$L asymmetry of the dark matter sector through the above interaction as long as $\Lambda$ satisfies $\Lambda \lesssim T_{\rm lept} (M_{\rm pl}/T_{\rm lept})^{1/(2n)}$. The last condition stems from the fact that the ADM interaction should be active after leptogenesis. Here, $T_{\rm lept}$ is the temperature in which leptogenesis occurs and is given by the decay temperature of right-handed neutrinos, and $M_{\rm pl} \simeq 2.43 \times 10^{18}$ GeV denotes the Planck scale. Interestingly, it is possible to predict the ratio of these asymmetries without knowing the details of the interaction as we shall see below. 

In our work we explicitly assume that the above ADM interaction decouples before the decoupling of the sphaleron processes. In general, asymmetry transfer may be active even when the ADM is non-relativistic. In such a case, the number density of the ADM is Boltzmann suppressed and a heavier ADM is required to explain the observed dark matter density \cite{heavy_ADM}. The mass of the ADM crucially depends on the nature 
of the interaction and  a wide mass range (up to TeV) may be realized.
The decoupling temperature of the ADM interactions is denoted by $T_D$, while the sphaleron decoupling is estimated to occur at $T_{sph} \simeq 137$ GeV~\cite{Sphaleron}. When the temperature of the universe is below $T_D$, both dark matter asymmetry and B$-$L asymmetry of the SM sector are individually conserved. In addition, when the temperature becomes lower than $T_{sph}$, all of dark matter asymmetry, B and L asymmetries of the SM sector are individually conserved. Using the relations among the chemical potentials of SM particles and the ADM particle at a temperature around $T_D$ (which are obtained from the SM and the ADM interactions), and the condition of neutrality of the universe, the ratio of B$-$L asymmetry in the SM and the dark matter sectors can be expressed as follows~\cite{IMY}:
\begin{eqnarray}
\frac{({\rm B} - {\rm L})_{\rm SM}}{\rm ({\rm B} - {\rm L})_{\rm DM}}
=
\frac{79}{22 \, Q_{\rm DM}^2}.
\end{eqnarray}
The above ratio does not depend upon the details of the ADM interaction in equation~(\ref{eq: ADM interaction}). Note also that the existence of a new singlet scalar does not alter the above relation as an SM singlet particle does not have any chemical potential.

The asymmetry $({\rm B} - {\rm L})_{\rm SM}$ is divided into baryon and lepton asymmetries of the SM sector (B$_{\rm SM}$ and L$_{\rm SM}$) when the sphaleron process decouples, where B$_{\rm SM}$ is the baryon asymmetry observed today. The ratio between $({\rm B} - {\rm L})_{\rm SM}$ and B$_{\rm SM}$ is given by ${\rm B}_{\rm SM}/({\rm B} - {\rm L})_{\rm SM} = 30/97$, which finally gives the ratio ${\rm B}_{\rm SM}/({\rm B} - {\rm L})_{\rm DM} = (30/97) (79/22) (1/Q_{\rm DM}^2)$. When the annihilation cross section between dark and anti-dark matter particles is large enough to eliminate its symmetric component in the early universe, $({\rm B} - {\rm L})_{\rm DM}$ is directly related to the dark matter abundance observed today, as in the case of baryon asymmetry. Asymmetries $({\rm B} - {\rm L})_{\rm DM}$ and ${\rm B}_{\rm SM}$ are then given by B$_{\rm SM} = \Omega_b \rho_c/(s_0 m_N)$ and $({\rm B} - {\rm L})_{\rm DM} = \Omega_{\rm DM} Q_{\rm DM} \rho_c/(s_0 m_{\rm DM})$. Here, the critical energy and entropy densities of the present universe are given by $\rho_c h^2 \simeq 1.05 \times 10^{-5}$ GeV/cm$^3$ and $s_0 \simeq 2890/$cm$^3$, and the dark matter and baryon abundance are given by $\Omega_{\rm DM} h^2 \simeq 0.120$ and $\Omega_b h^2 \simeq 0.0220$~\cite{Planck}, where $h \simeq 0.670$ is the scale factor for the Hubble expansion rate. The ADM mass is denoted by $m_{\rm DM}$ and the nucleon mass $m_N \simeq 938$ MeV. Using the relation between $({\rm B} - {\rm L})_{\rm DM}$ and ${\rm B}_{\rm SM}$, the mass of the ADM particle is then found to be
\begin{eqnarray}
m_{\rm DM}
= \frac{30}{97} \frac{79}{22}
\frac{\Omega_{\rm DM}}{\Omega_b} \frac{m_N}{Q_{\rm DM}}
\simeq \frac{5.7 \, {\rm GeV}}{Q_{\rm DM}}.
\label{eq: DM mass}
\end{eqnarray}

\subsection{Fermionic ADM with a light scalar mediator}
\label{subsec: lag}

In the previous subsection, we briefly reviewed a simple scenario for light ADM with a B$-$L asymmetry, where, under certain mild assumptions, the mass of the DM particle is predicted as a function of its B$-$L charge (equation~(\ref{eq: DM mass})). The only interaction between the DM sector and the SM sector assumed for this scenario so far is the total B$-$L number conserving but DM number violating interaction given by equation~(\ref{eq: ADM interaction}). This interaction does not lead to any annihilation between DM and anti-DM particles. The lowest dimension effective interaction connecting the DM particles to the SM sector is described by the following dimension-5 operator
\begin{equation}
\mathcal{L}_5=\frac{\lambda}{\Lambda}|H|^2 (\overline{\chi}\chi+{\rm h.c.}),
\end{equation}
where $H$ denotes the SM Higgs doublet and $\chi$ is the fermionic DM particle. However, since the annihilation cross section via this interaction is suppressed by the cutoff scale $\Lambda$, it is difficult to eliminate the symmetric component with this term alone when $\Lambda > {\cal O}(1)$ TeV. We, therefore, do not consider this possibility any further. Instead, we introduce an additional light state which couples to the fermionic DM particle. This additional light state can either be spin-0 or spin-1. Such an interaction can also play the role of connecting the DM sector with the SM sector. In the scalar mediator case, the mediator can mix with the SM Higgs boson giving rise to a Higgs-portal interaction, and in the vector mediator case, one can have kinetic mixing with the SM gauge sector. Since the scalar mediator case does not require the introduction of new gauge interactions, we focus on this possibility only. The vector mediator option has been discussed in detail in reference~\cite{Zurek}. The Lagrangian describing the DM and the light scalar sectors is given by
\begin{eqnarray}
&& \mathcal{L} =
\textit{i}~\overline{\chi}(\dslash-m_{\chi})\chi
+\frac{1}{2}(\partial_\mu \phi^\prime \partial^\mu \phi^\prime
-m_{\phi^\prime}^2 {\phi^\prime}^2)
-\kappa \overline{\chi}\chi \phi^\prime
-V(H^\prime,\phi^\prime),
\label{eq-Lag} \\
&& V(H^\prime,\phi^\prime) =
V(H^\prime)_{SM}
+\lambda_1 \phi^\prime |H^\prime|^2
+\lambda_2 {\phi^\prime}^2 |H^\prime|^2
+\lambda_3 {\phi^\prime}^3
+\lambda_4 {\phi^\prime}^4.
\label{Eq: Lag}
\end{eqnarray}
Here, $V(H^\prime)_{SM}$ represents the usual SM Higgs potential\footnote{For a study on the vacuum stability of such a scalar potential, see, for example, reference~\cite{Ko-2}.}. Without any loss of generality, we assume that the SM singlet scalar field $\phi^\prime$ does not receive any vacuum expectation value (VEV), since the VEV of a singlet scalar field can be absorbed by re-defining the field and its associated coupling constants. After electroweak symmetry breaking, the neutral component of the SM Higgs doublet, $h^{\prime}$, will mix with $\phi^\prime$. We can thus express the mass eigenstates as follows:
\begin{eqnarray}
h &=& (\cos\alpha) \, h^{\prime} - (\sin\alpha) \, \phi^{\prime},
\nonumber \\
\phi &=& (\sin\alpha) \, h^{\prime} + (\cos\alpha) \, \phi^{\prime}.
\end{eqnarray}
Clearly, for small values of mixing angles, the mass eigenstate $h$ is mostly SM-like, and $\phi$ is mostly singlet-like. Due to this mixing, the couplings of the SM-like Higgs boson to all fermions and gauge bosons get modified by the factor $\cos\alpha$, while the particle $\phi$ now couples to the SM gauge bosons and fermions, with a strength proportional to $\sin\alpha$. Thus, we have the following relations,
\begin{eqnarray}
g(h X \bar{X}) &=& (\cos\alpha) \, g(h X \overline{X})_{SM} \nonumber \\
g(\phi X \bar{X}) &=& (\sin\alpha) \, g(h X \overline{X})_{SM},
\end{eqnarray} 
where $X$ represents any SM gauge boson or fermion, and $g(h X \overline{X})_{SM}$ denotes the corresponding coupling strength between the SM Higgs boson and $X$.

The phenomenology for the parameter region $m_\phi>m_\chi$ has been studied in reference~\cite{Russell}, and we do not discuss it in this paper. The region $m_{\phi} \simeq 2m_{\chi}$ leads to a resonant enhancement of dark matter pair annihilation to an SM fermion pair~\cite{Russell}. Although the resonance region can satisfy the relic density requirement and is much less constrained from direct detection experiments, it is somewhat fine-tuned. In this study, we focus on the $m_{\phi} < m_{\chi}$ case, and explore its phenomenology in detail. For the DM charge under $U(1)_{\rm B-L}$, we take two examples: $Q_{\rm DM} = 1/3$ or 5/3. This leads to the DM mass $m_\chi$ of 17.1 GeV and 3.42 GeV, respectively.

%%%%%%%%%%%%%%%%%%%%%%%%%%%%%%%%%%%%%%%%%%
\begin{figure}[t]
\begin{center}
\centering\includegraphics[angle=0, width=10.5cm]{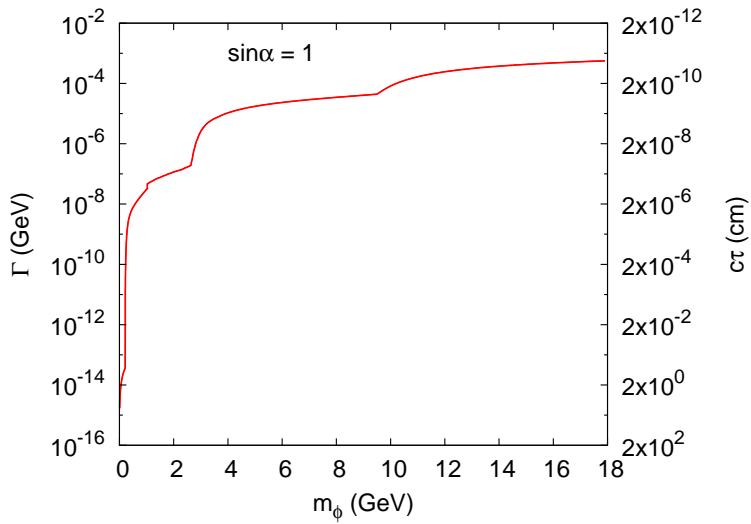}
\caption{\small \sl Decay width and decay length of the scalar mediator $\phi$, with $\sin \alpha = 1$.}
\label{fig: lifetime}
\end{center}
\end{figure}
%%%%%%%%%%%%%%%%%%%%%%%%%%%%%%%%%%%%%%%%%

The lifetime and decay branching ratios of $\phi$ ($m_\phi < 17.1$ GeV for the above choice of DM masses) is important in the phenomenology of this model, both at colliders and also in cosmology. Since the couplings of $\phi$ to the SM particles are suppressed by the factor $\sin^2 \alpha$, its decay width is smaller than that of an SM Higgs boson of the same mass. For $m_{\phi}< 2m_e$, the mediator can only decay to a pair of photons via top and $W$ loops. In the mass range $2 m_e < m_{\phi} < 2 m_{\mu}$, $\phi$ decays almost entirely to electrons. Below the $2 m_{\pi}$ threshold, it can decay dominantly to muons, while above it, decays to pion pairs are also possible through its interaction with gluons. In the mass range of 1-2 GeV, further decay channels to hadronic final states open up. For $m_\phi > 3$ GeV, $\phi$ can decay to charm quarks or $\tau$ leptons. Once the $b \bar{b}$ mode opens up, it dominates $\phi$ decay with a branching ratio of about 85\%.

We have used the code HDECAY~\cite{Djouadi:1997yw} for the calculation of decay widths and branching ratios of $\phi$ for $m_\phi > 3$ GeV, by scaling the SM Higgs boson widths of the same mass appropriately. Below the charm threshold, $\phi$ decay to hadrons are important, but these cannot be calculated with HDECAY. For this reason, for $\phi$ masses up to 3 GeV, following reference~\cite{Gunion:1989we}, we have used the widths which can be obtained by assuming a phenomenological spectator approximation valid above the two pion threshold. We show the decay width ($\Gamma$) and the decay length ($c \tau$) of the mediator as a function of its mass in figure~\ref{fig: lifetime} (for $\sin\alpha = 1$). Due to the use of approximate methods, in the range $2 m_{\pi}< m_\phi<3$ GeV, the total width and the branching ratios to $\mu^{+} \mu^{-}$ or $e^{+} e^{-}$ channels have some uncertainty.

For very small mediator masses of around 100 MeV, $\phi$ can be long-lived, and the lifetime increases as the mixing angle becomes smaller (for very small mixing angles, even a higher mass $\phi$ will be long-lived). There can also be an additional boost factor depending upon the production mechanism at colliders, which may increase the decay length further. Therefore,  this fact might facilitate the collider search for $\phi$ at the LHC.

%%%%%%%%%%%%%%%%%%%%%%%
%%%%% Constraints %%%%%
%%%%%%%%%%%%%%%%%%%%%%%
\section{Current constraints}
\label{sec: constraints}

%%%%%%%%%%%%%%%%%%%%%%%%%%%%%%%%%%%%%%%%%%
\begin{figure}[p]
\begin{center}
\centering\includegraphics[angle=0, width=14cm]{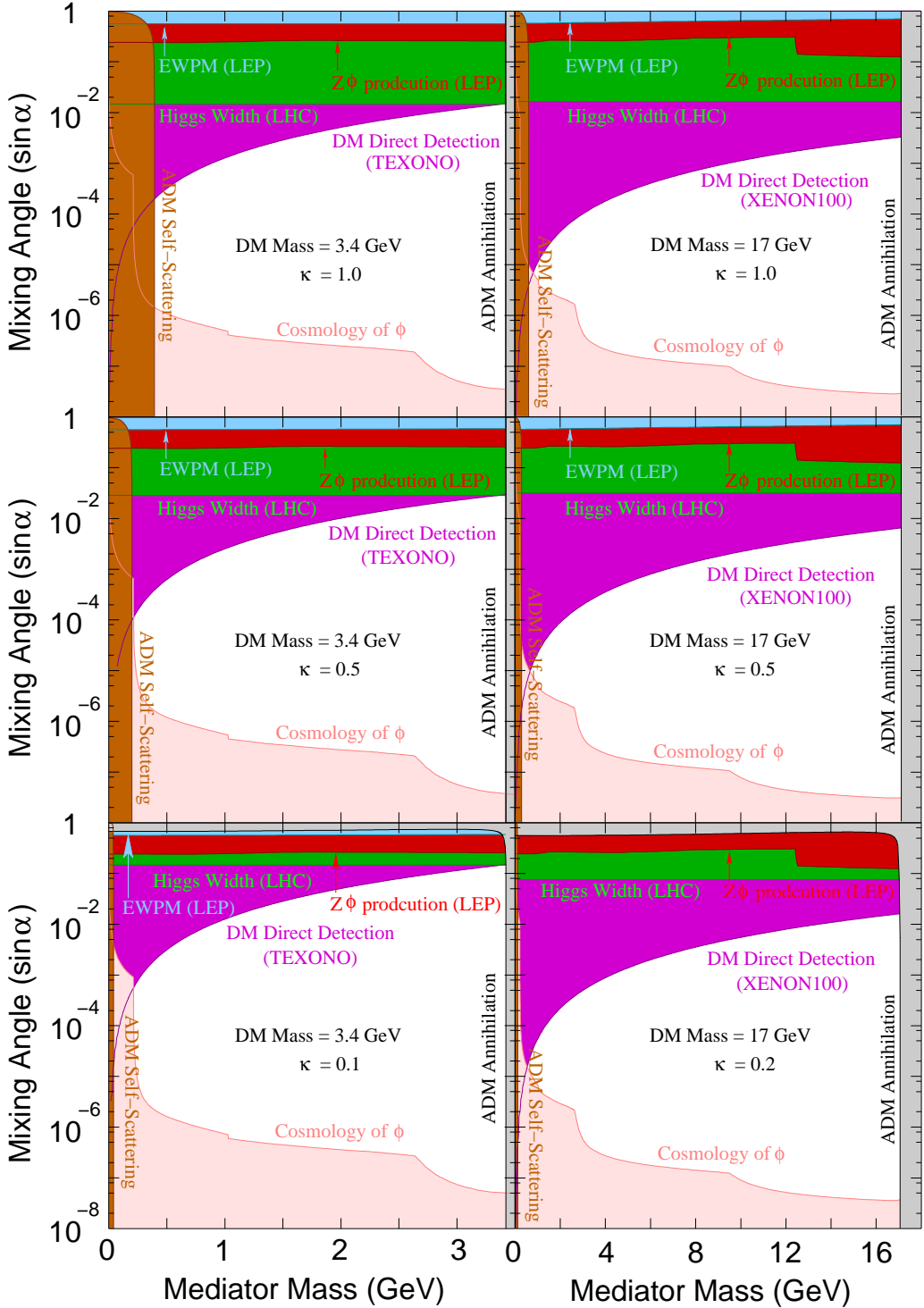}
\caption{\small \sl Current constraints on the ($m_\phi$, $\sin\alpha$)-plane from cosmology (annihilation/self-scattering of ADM and $\phi$-cosmology), DM direct detection experiments (XENON100 and TEXONO), the LEP search for light scalars, precision electroweak data and the global fit upper bound on the invisible branching ratio of the Higgs boson from LHC data.}
\label{fig: P}
\end{center}
\end{figure}
%%%%%%%%%%%%%%%%%%%%%%%%%%%%%%%%%%%%%%%%%

In this section, we discuss in detail the constraints on the GeV scale ADM model parameter space coming from cosmological observations, dark matter direct detection experiments (XENON100~\cite{Xenon100} and TEXONO~\cite{Texono}), collider experiments at the LEP and LHC, and flavour physics constraints coming from BaBar and CLEO.

\subsection{Constraints from cosmology}

\subsubsection{Relic abundance}

As argued before, as long as the annihilation cross section of dark matter particles is large enough to eliminate the symmetric component in the early universe, the present relic density is determined by the asymmetry itself. Since we are concentrating on the parameter region where $m_\phi < m_\chi$, the most important annihilation process is $\chi \overline{\chi} \rightarrow \phi \phi$ through t-channel $\chi$ exchange. The annihilation process $\chi \overline{\chi} \rightarrow \phi \phi$ can also proceed through s-channel $\phi$ exchange, if the $\phi^3$ self-coupling is large enough (presumably, at the same time, being less than the quartic $\phi$ coupling, in order not to de-stabilize the vacuum). We neglect this possibility in our analysis. Furthermore, the annihilation into SM fermions, $\chi \overline{\chi} \rightarrow f \bar{f}$ , which can proceed via s-channel $\phi$ or $h$ exchange are not important except in the resonance region $m_\phi \simeq 2m_\chi$. Therefore, we can safely ignore them in the parameter region of our interest.

The relic density calculation for ADM scenarios has been presented in several references~\cite{Zurek,Russell,Drees}. Since the parameter space $m_\phi < m_\chi$ is away from any resonance or threshold regions, we can use an expansion of the annihilation cross section $\sigma$ in terms of the dark matter relative velocity $v$. Considering the t-channel and u-channel $\chi$ exchange diagrams for the process $\chi \overline{\chi} \rightarrow \phi \phi$, we obtain
\begin{equation}
\sigma v =
\frac{\kappa^4 \cos^4\alpha (s-4m_\chi^2)}{24 \pi (2m_\chi^2-m_\phi^2)^4}
(9m_\chi^4-8m_\chi^2 m_\phi^2+2m_\phi^4) \sqrt{1-\frac{m_\phi^2}{m_\chi^2}},
\label{Eq: sigma-v}
\end{equation}
where, $s$ is the centre of mass energy. The thermally averaged annihilation cross section $\langle \sigma v \rangle_x$ at a reduced inverse temperature $x=m_\chi/T$ can be approximated using equation~(\ref{Eq: sigma-v}) as follows:
\begin{equation}
\langle \sigma v \rangle_x \simeq 3 \times 10^3 ~\text{pb} \left(\frac {10 ~\text{GeV}}{m_\chi}\right)^2 \left(\frac{30}{x}\right) \kappa^4 \cos^4\alpha.
\label{Eq: relic}
\end{equation}
The decoupling temperature for a thermal relic is generically found to be $x_F \simeq$ 20--30. Using equation (\ref{Eq: relic}), we can put a lower bound on the coupling strength $\kappa$ of the DM particles with the mediator $\phi$. This lower bound is obtained by demanding that $\langle \sigma v \rangle_{x_F} > 1~{\rm pb}$, such that the symmetric component is almost completely annihilated before decoupling. Using this condition, for $m_\chi=17.1$ GeV, we find that $\kappa >$ 0.16--0.18, where the range of values is obtained by varying $x_F$ in the range $20-30$. Thus, in this case, we take the conservative estimate of $\kappa \geq 0.2$. For $m_\chi=3.42$ GeV, a similar estimate gives us $\kappa \geq 0.1$. The lower bound on $\kappa$ for a given DM mass is the most important constraint coming from the relic density requirement. In figure~\ref{fig: P}, we show by grey bands the regions which are disallowed by the requirement of a minimum value of the annihilation cross section as a function of $m_\phi$ and $\sin \alpha$. Clearly, the above estimate receives only small corrections from the exact numerical evaluation of the annihilation rate, and almost the entire regions are allowed as long as we satisfy the above lower bounds on $\kappa$.

\subsubsection{DM Direct Detection} 

In this subsection, we derive the constraints on the ADM scenario from dark matter direct detection experiments. Dark matter, present in the local halo, can occasionally interact with the nucleons inside a detector. The dominant interaction mechanism is elastic spin-independent scattering, and the direct detection experiments aim to measure the nucleon recoil energy. Typically, the energy of the scattered nucleus, induced by interaction with dark matter, varies from several eV to hundreds of keV depending on the mass of both the dark matter ($m_{\chi}$) and the nucleus. The relevant quantity, measured in the direct detection experiments, is the differential event rate, defined as the number of events per day run time per keV recoil energy per kg detector mass. The event rate can be calculated by integrating this differential rate over a particular energy range, determined by the nature of the detector material. The differential rate is a function of the dark matter flux, velocity distribution, density of the target nuclei and dark matter-nucleon scattering cross section ($\sigma$). The dark matter-nucleon cross section can be expressed in terms of dark matter-quark effective interactions, which depend on the underlying particle physics model. For the ADM model under study, the primary processes responsible for spin-independent scattering are through t-channel exchange of $\phi$ or $H$ between the ADM particles and quarks or gluons, in the latter case via heavy quark loops.

The spin-independent scattering cross section between the dark matter and a nucleon in the ADM model is dominated by the process through t-channel exchange of $\phi$ when $m_\phi << m_h$, which is given by the following equation
\begin{equation}
\sigma_{SI} =
\frac{(\cos^2 \alpha \, \sin^2\alpha) \kappa^2}{\pi v^2 m^4_{\phi}}
\frac{m^2_p m^2_{\chi}}{(m_{\chi} + m_p)^2} f^2_p
\label{sigmaSI}
\end{equation}
with $f_p/m_p = f_{Tu} + f_{Td} + f_{Ts} + (2/9) f_{TG}$. Here, $v \simeq 246$ GeV is the VEV of the SM Higgs doublet and $m_p$ denotes the nucleon mass. The quantity $f_p$ is determined by several hadronic matrix elements and the first three terms are the contributions from light quarks to the composition of the nucleon, while the sum of heavy quark contributions (or equivalently gluon contribution) is given by $f_{TG}$. We take $f_{Tu} = f_{Td} \simeq 0.028$ and $f_{Ts} \simeq 0$~\cite{Ohki:2009mt} for our numerical calculations. The last term is determined as $f_{TG} \simeq 0.943$, through the trace anomaly relation, $f_{Tu} + f_{Td} + f_{Ts} + f_{TG} = 1$~\cite{Kanemura:2010sh}. Since the scattering cross section is found to be inversely proportional to the fourth power of the mass of the exchanged scalar, the Higgs boson contribution can be neglected compared to that of the light mediator.

There are a number of DM direct detection experiments that are currently running or are under development. For high mass dark matter, XENON100~\cite{Xenon100}, a liquid Xenon detector puts the strongest bound on the spin-independent cross section to date. The upper limit on $\sigma_{SI}$ for $m_{\chi} \sim $17 GeV is about $2 \times 10^{-44}$ cm$^2$ at the 90\% C.L., which can be used to calculate the bound on the 17.1 GeV ADM case. Since the recoil energy of the nucleon becomes smaller for light dark matter, it becomes very challenging for direct detection experiments, as the construction of detectors with a very low threshold is difficult. This makes the XENON100 experiment lose its sensitivity for dark matter masses below 5 GeV, due to its high threshold. In such cases, XENON10~\cite{Xenon10}, CRESST~\cite{Cresst} and TEXONO~\cite{Texono} experiments are the most sensitive probes, among which we find that the best limit comes from the TEXONO experiment. The constraints from XENON100 for $m_\chi = 17.1$ and from TEXONO for $m_\chi = 3.4$ GeV on the ($m_\phi$, $\sin\alpha$)-plane are shown in figure~\ref{fig: P} for different allowed values of dark matter-mediator coupling $\kappa$ (the pink shaded regions are disallowed by the current data). We note that, in general, the bounds obtained from the XENON100 or TEXONO experiments are much stronger than other constraints irrespective of the mass of the mediator and $\kappa$. As we can see from this figure, the strongest constraints are for lower values of $m_\phi$, and $\sin \alpha$ has an upper bound of around $10^{-2}$ to $10^{-6}$ depending upon the $\phi$ mass. 

\subsubsection{Other cosmological constraints}

As in the case of scattering cross section between dark matter and nucleon, the existence of a light scalar mediator also enhances the self-scattering cross section between the dark matter particles, which may affect DM halo dynamics significantly. The self-scattering cross section is indeed constrained by several astrophysical observations, for example, observations of bullet clusters, elliptical galaxy clusters, and elliptical dark matter halos. The most stringent limit comes from dark matter halos, because the self-scattering of dark matter particles leads to spherical DM halos and thus the observed ellipticity of DM halos puts an upper limit on the cross section. According to reference~\cite{Zurek}, we can put a limit on the model parameters using $\sigma_T < 4.4 \times 10^{-27} \, {\rm cm}^2 \, (m_\chi/1 \, {\rm GeV})$. Here, $\sigma_T$ is the self-scattering cross section weighted by the momentum transfer, and its explicit form is given by
\begin{eqnarray}
\sigma_T = \frac{\cos^4\alpha \, \kappa^4}{4\pi} \frac{m_\chi^2}{m_\phi^4}.
\end{eqnarray}
This constraint is shown in figure~\ref{fig: P}, where, the brown shaded regions are disallowed by this limit. As we can see in this figure, the self-scattering constraint leads to a lower limit on the mediator mass $m_\phi$, when the mixing angle is small and $\kappa$ is large.

On the other hand, a lower limit on the mixing angle $\sin\alpha$ is obtained by considering the thermal history of the light scalar mediator $\phi$. Firstly, the lifetime of the mediator $\phi$ is longer when the mixing angle is smaller. The lifetime must be, however, shorter than 1 (10$^{-2}$) sec when $\phi$ mainly decays leptonically (hadronically), in order not to spoil the successful big-bang nucleosynthesis. The limit on the angle then becomes $\sin\alpha > 10^{-9}$ for $m_\phi > 2 m_\pi$, $\sin\alpha > 10^{-8}$ for $2 m_\pi > m_\phi > 2 m_\mu$, and $\sin\alpha > 10^{-6}$ for $2 m_\mu > m_\phi > 2 m_e$. Furthermore, a stronger limit can be obtained when we impose the condition that the dark matter particles are in thermal and chemical equilibrium with the SM particles at the freeze-out temperature (the temperature in which the symmetric component of the ADM particles is eliminated), as in the case of usual cosmology. A lower limit on the mixing can be derived by considering the ratio of the reaction rate maintaining the equilibrium ($\Gamma$) to the expansion rate of the universe (the Hubble parameter $H$) at the freeze-out temperature. Here, $\Gamma$ is given by
\begin{eqnarray}
\Gamma = \frac{1}{\tau} \frac{K_1(m_\phi/T_f)}{K_2(m_\phi/T_f)},
\end{eqnarray}
with $\tau$, $T_f$, and $K_n(z)$ being the lifetime of $\phi$, the freeze-out temperature, and the modified Bessel function of the second kind of degree $n$, respectively. The dark matter particles cannot be in equilibrium with SM particles when $\Gamma/H < 1$, as in the light-pink shaded parameter region shown in figure~\ref{fig: P}. It can be seen from this figure that the lower limit on the mixing angle $\sin\alpha$ turns out to be as small as 10$^{-6}$--10$^{-7}$, unless $m_\phi$ is too small.

\subsection{Collider constraints}

\subsubsection{LEP limits}

The LEP experiments searched for light scalar particles with SM Higgs-like couplings to gauge bosons and fermions in the $e^+ e^- \rightarrow Z \phi$ channel, where the $\phi$ mass could be determined by using the recoil of the Z boson decay products and the initial known center of mass energy. The absence of any signal led to bounds on the ratio
\begin{equation}
\sin^2\alpha = \left(\frac{g(\phi ZZ)}{g(hZZ)_{SM}}\right)^2.
\end{equation}
The OPAL collaboration reported $95 \%$ C.L. upper bounds on $\sin^2\alpha$ independent of the decay mode of $\phi$ from the recoil mass spectrum of $Z \rightarrow e^+e^-/\mu^+ \mu^-$~\cite{OPAL}. Furthermore, LEP combined results were presented for $m_\phi \gtrsim 5$ GeV using the $\phi \rightarrow \tau^+ \tau^-$ decay mode, and for $m_\phi\gtrsim 12.5$ GeV using the $b \bar{b}$ mode~\cite{LEP-ZH}. While the $\tau^+ \tau^-$ decay mode did not achieve much improvement over the decay mode independent OPAL limit, the $b \bar{b}$ decay mode improves the bounds by roughly one order of magnitude. We combine the limits from LEP by taking the most stringent bound for a given $m_\phi$, and show them in figure~\ref{fig: P} as the LEP $Z\phi$ lines (with the red shaded regions being ruled out) on the ($m_\phi$, $\sin\alpha$)-plane. One can approximately express the bounds in the $m_\phi$ range of our interest as follows:
\begin{equation}
\sin^2\alpha \lesssim
\begin{cases}
\, 0.10 \qquad \left( \, 1.00 \, {\rm keV}\, < m_\phi < 12.5 \,{\rm GeV} \, \right),
\\
\, 0.02 \qquad \left( \, 12.5 \, {\rm GeV}\, < m_\phi < 17.1 \,{\rm GeV} \, \right).
\end{cases}
\end{equation}

Due to the mixing of the singlet scalar with the Higgs, the oblique parameters $S$ and $T$ get modified (from self-energy diagrams involving the $W$ or $Z$ bosons in this case), and therefore, the LEP electroweak precision data puts another constraint on the mixing angle. However, we find that, for the mass range of $\phi$ we are interested in, the bounds from the precision data are rather weak compared to the constraints from the LEP $Z\phi$ search channel. In scenarios with mixed-in singlets, several references have computed the upper bounds on $\sin \alpha$~\cite{Ko,Wells}, mostly concentrating on a singlet-like eigenstate with much higher mass. After calculating the contributions to the $S-T$ parameters following the one-loop expressions as given in reference~\cite{Ko}, and comparing them with the experimentally allowed $95\%$ confidence level range in the $\Delta S$-$\Delta T$ plane, as described in reference~\cite{Beringer:1900zz}, we obtain (taking $m_h=126$ GeV) the following bound on the mixing angle:
\begin{equation}
\sin^2\alpha < 0.8 \qquad \left( \, 95\% \,{\rm C.L.} \, \right).
\end{equation}
As we can see, this bound is weaker than the one from the LEP direct search for a light scalar in the $Z\phi$ channel, in the entire range of $m_\phi$ considered by us.

\subsubsection{Bounds from LHC Higgs data}
\label{sec:LHC-Higgs}

We have seen in section~\ref{subsec: lag} that there are two important modifications to the SM-like Higgs boson properties when compared with the standard model. Firstly, due to its mixing with the singlet, all the couplings of the SM-like Higgs boson get scaled by the factor $\cos\alpha$, leading to an universal suppression of the signal strengths measured in different channels at the LHC. Secondly, the total width of the Higgs boson receives additional contributions from both the invisible decay mode $h \rightarrow \chi \overline{\chi}$ as well as the decay $h \rightarrow \phi \phi$. The latter decay, however, depends upon additional parameters in the model, for e.g., $\lambda_2$ in equation~(\ref{Eq: Lag}). To simplify the analysis, we have parameterized this in terms of the width $\Gamma_{h\rightarrow \phi \phi}$ itself, which also plays a major role in the direct collider search for the $\phi$ particle, as will be discussed in the next sub-section. The constraints expected in the Higgs portal DM models from the LHC Higgs data have been discussed in previous studies~\cite{Kanemura:2010sh, Ko, Higgs portal}. In our scenario, since the invisible width and the reduction in the signal strengths in the visible channels are both functions of the mixing, we perform a global analysis of the Higgs data to determine the allowed ranges for $\sin^2\alpha$ as well as $\Gamma_{h\rightarrow \phi \phi}$, for our chosen values of $\kappa$. The dataset used for this analysis includes 10 data points from ATLAS and CMS Higgs measurements, namely, the signal strengths, from each experiment, in the $\gamma \gamma$, $ZZ^*$, $WW^*$, $\tau^+ \tau^-$ and $b \bar{b}$ modes. We use the combination of 7 TeV (about 5 fb$^{-1}$) and 8 TeV (about 20 fb$^{-1}$) results, details of which can be found in references~\cite{ATLAS-h, CMS-h}. The method of analysis used here is described in reference~\cite{BMM}. In our scenario, the signal strength in each Higgs search channel is given by
\begin{equation}
\mu =
\frac{\cos^4 \alpha \, \Gamma_h^{SM}}
{\cos^2 \alpha \, \Gamma_h^{SM}
+\Gamma_{h \to \chi \bar{\chi}}+\Gamma_{h \to \phi \phi}},
\end{equation}
where, the invisible Higgs decay width $\Gamma_{h \rightarrow \chi \overline{\chi}}$ is 
\begin{equation}
\Gamma_{h \rightarrow \chi \overline{\chi}}=\frac{\kappa^2 \sin^2{\alpha}}{8 \pi} m_h \left(1-\frac{4m_{\chi}^2}{m_h^2}\right)^{3/2}.
\end{equation}
We therefore have two parameters ($\sin^2\alpha$ and $\Gamma_{h\rightarrow \phi \phi}$) in our global fit, where the latter has been varied in the range $[0 : 2\Gamma_h^{SM}]$, $\Gamma_h^{SM} =$ 4.21 MeV being the total decay width of an SM Higgs boson of mass 126 GeV~\cite{LHCWG}. While determining the $95 \%$ C.L. upper bound on one parameter, we have marginalized over the other one.

The fit yields a $\chi^2_{\rm min}$ of 9.93, for 8 degrees of freedom ($\chi^2_{\rm min}$/d.o.f$=1.24$). At the global minimum of $\chi^2$, both $\sin^2 \alpha$ and $\Gamma_{h \rightarrow \phi \phi}$ take the value $0$, reflecting the fact that LHC Higgs data does not leave much room for deviations from the SM predictions, especially in a scenario where the signal strengths in all channels are universally suppressed. We can express the $95\%$ C.L. upper limits on $\sin^2\alpha$ (after marginalizing over $\Gamma_{h\rightarrow \phi \phi}$), for different values of $m_\chi$ and $\kappa$ as follows:
\begin{equation}
\sin^2\alpha <
\begin{cases}
\, 2.4 \, \kappa^{-2} \times 10^{-4} \qquad
\left( \, m_\chi=17.1 \, {\rm GeV} \, \right),
\\
\, 2.1 \, \kappa^{-2} \times 10^{-4} \qquad
\left( \, m_\chi=3.4 \, {\rm GeV} \, \right).
\end{cases}
\label{HiggsLHC}
\end{equation}
One can also translate the above bounds to put an upper limit on the allowed invisible branching ratio of the Higgs boson, which is given by
\begin{equation}
{\rm BR}(h \rightarrow \chi \overline{\chi}) < 20\%
\qquad
(95\% \, {\rm C.L., \, marginalized \, over} \, \Gamma_{h \to \phi \phi}).
\end{equation}
Our results on the invisible branching ratio are in agreement with the results obtained, for example, in references~\cite{InvisibleH}. For such low allowed values of $\sin^2\alpha$, the correlated variation of the visible modes and the invisible width (both being functions of $\sin^2\alpha$) does not play any major role. The present data, therefore, essentially puts an upper bound on the total non-standard branching ratio of the Higgs. In our case, if we now marginalize over $\sin^2\alpha$ instead of $\Gamma_{h \to \phi \phi}$, we also find an exactly same upper bound of $20 \%$ (at $95 \%$ C.L.) on ${\rm BR} ({h\rightarrow \phi \phi})$.

The bounds on $\sin^2\alpha$ in equation~(\ref{HiggsLHC}) indicate that the LHC Higgs data already puts quite severe constraints on our ADM scenario. In particular, as we can see from figure~\ref{fig: P}, it excludes substantial regions in the parameter space for higher values of $\kappa$ (indicated by the green shaded regions). Moreover, for low ADM mass, where the direct detection experiments have lower sensitivity, the bounds from the LHC Higgs data remain an important independent probe.

\subsubsection{LHC light scalar search} 
\label{sec:LHC-mediator}

At the LHC, the most important production mechanisms of $\phi$ are similar to that of the SM Higgs boson, viz., gluon fusion, weak-boson fusion and associated production with W/Z or top quarks. Additionally, the mediator may be produced from Higgs boson decays. Direct $\phi$ production cross section is determined by $\sin\alpha$, whereas in case of Higgs decays to mediator, the branching ratio also depends on the parameter $\lambda_2$ (see equation~\ref{Eq: Lag}), which is otherwise unconstrained and may make the rate substantial. Thus in general, the mediator search strategy can be divided into two classes: (a) direct $\phi$ production and (b) $\phi$ production from Higgs decays, where the latter depends on the size of the Higgs branching to a $\phi$ pair. As already discussed, the possible mediator decay modes are determined by its mass. \\

\begin{figure}[t]
\begin{center}
\centering\includegraphics[angle=0, width=10.5cm]{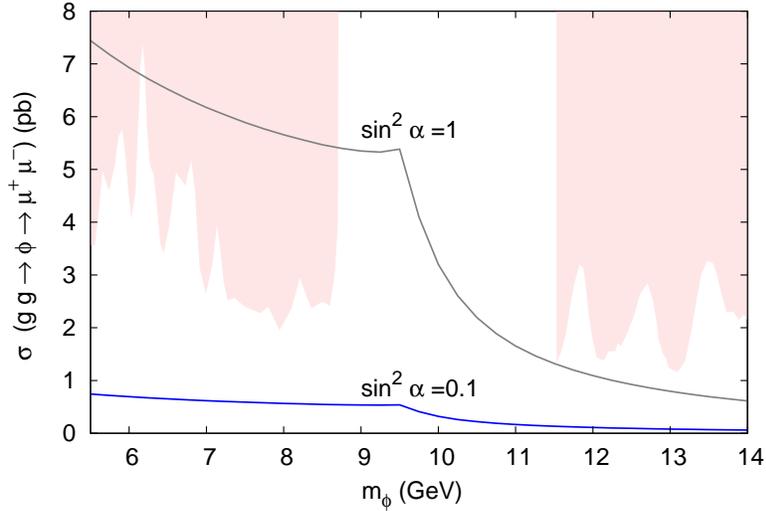}
\caption{\small \sl $\sigma(g g \rightarrow \phi) \times {\rm Br}~(\phi \rightarrow \mu \mu$) as a function of the mediator mass. The shaded region is excluded by the CMS di-muon resonance search at 7 TeV LHC (with 1.3 fb$^{-1}$ data). The black and blue curves correspond to $\sin^2 \alpha$ = 1 and 0.1 respectively.}
\label{fig:phimumu}
\end{center}
\end{figure}

{\bf Direct production:} \, The dominant $\phi$ production is from the gluon fusion process, in the entire mass range considered in our study. Although the cross section can be large enough (for example, at the 7 TeV LHC, $\sigma(g g \rightarrow \phi) \sim$ 400 pb for $m_{\phi}=5$ GeV, assuming $\sin^2 \alpha =0.1$), detection of such light particles is not an easy task because of huge SM backgrounds in the kinematic region of interest. If $\phi$ dominantly decays to quarks, the gluon fusion process cannot be used, since this signal will be overwhelmed by the QCD two-jet background. The only channels which can be exploited are $g g \rightarrow \phi \rightarrow \mu^+ \mu^-$ or $\tau^+ \tau^-$. At the 7 TeV run, the CMS collaboration has looked for a narrow resonance in the opposite sign di-muon invariant mass distribution using 1.3 fb$^{-1}$ of data~\cite{CMS2mu}. Although, this analysis was designed to look for a light pseudo-scalar boson, in the mass range of 5.5 to 14 GeV (in the context of the next to minimal supersymmetric standard model), this limit can be translated to constrain the mediator production in the ADM model. For this purpose, we have computed the $g g \rightarrow \phi$ production cross section at NNLO using the code HIGLU \cite{Spira:1995mt}. We have used the MSTW2008NNLO~\cite{MSTW} parton distribution functions, with the factorization and renormalization scales fixed at the mass of the mediator. It should be noted that the mediator production cross section, for the small values of $m_\phi$ considered by us, has a large scale uncertainty (about 25\%). In figure~\ref{fig:phimumu}, we show $\sigma(g g \rightarrow \phi) \times {\rm Br}~(\phi \rightarrow \mu \mu$) as a function of $m_\phi$, for $\sin^2 \alpha$ = 1 and 0.1. The region excluded by the CMS di-muon resonance search is indicated by the shaded area. The CMS analysis is not sensitive enough in the 8-11 GeV mass window because of a large background from $\Upsilon$ states that can decay to muon pairs. This result shows that the current limit can only exclude a very large mixing scenario (i.e., $\sin^2 \alpha \sim $1), for $m_\phi \lesssim 9$ GeV . For $\phi$ masses above 9 GeV, the branching fraction Br($\phi \rightarrow \mu \mu $) is very small and the present sensitivity is not sufficient to probe most of the parameter space of the ADM model under consideration. \\

%%%%%%%%%%%%%%%%%%%%%%%%%%%%%%%%%%%%%%%%%%
%%%%%%%%%%%%%%%%%%%%%%%%%%%%%%%%%%%%%%%%%

%%%%%%%%%%%%%%%%%%%%%%%%%%%%%%%%%%%%%%%%%%
\begin{figure}[t]
\begin{center}
\centering\includegraphics[angle=0, width=10.5cm]{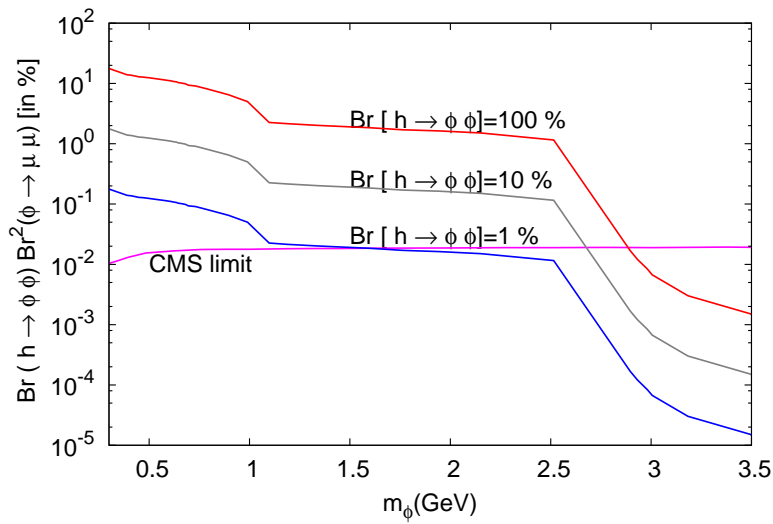}
\caption{\small \sl Constraints on the branching fraction Br(h $\rightarrow \phi \phi$) from CMS $4 \mu$ search, using 5.3 fb$^{-1}$ data at 7 TeV LHC. Contours of fixed BR ($h \rightarrow \phi \phi$) are shown as blue, black and red curves (1\%, 10\% and 100\%), while the CMS bound is shown as a pink line.}
\label{fig:4mu}
\end{center}
\end{figure}
%%%%%%%%%%%%%%%%%%%%%%%%%%%%%%%%%%%%%%%%%

{\bf Mediator from Higgs decay:} \, Since the mediator particle we consider is always much lighter than the Higgs boson, it can be produced from Higgs decays. Although the Br$(h \rightarrow \phi \phi)$ is {\it a priori} undetermined in our model, in section~\ref{sec:LHC-Higgs}, we found an upper limit of 20\% from a global fit of the present Higgs data. This gives us a promising opportunity to discover the light mediator particle from Higgs decays as long as the $h \rightarrow \phi \phi$ branching ratio is sizable. As before, the search prospects depend critically on the $\phi$ branching ratios to different final states. The CMS collaboration has searched for a new scalar particle ($a$) produced in Higgs decays, in the four muon channel, using 5.3 fb$^{-1}$ of data at the 7 TeV LHC~\cite{CMS4mu}. No significant excess has been found over SM backgrounds, which leads to an upper limit on $\sigma(p p \rightarrow h \rightarrow a a) \times \mbox{Br}^2(a \rightarrow \mu^+ \mu^-) $ as a function of the mass of $a$. We have used this result to evaluate the current limit on the Higgs branching ratio to a $\phi$ pair, and our results are shown in figure~\ref{fig:4mu}. Clearly, this limit is applicable only in the region $2m_e < m_{\phi} <2m_{c}$.  For  $\phi$ masses above the charm quark threshold,  final states involving tau leptons and bottom quarks become important, the analysis for which has not yet been reported by the LHC collaborations. As we can see from this figure, unlike in the case of direct $\phi$ production, the bound on Br(h $\rightarrow \phi \phi$ ) from LHC is already very strong. For $m_\phi < 1.6$ GeV, Br(h $\rightarrow \phi \phi$) is constrained to be lower than 1\%, while for $1.6 ~{\rm GeV} < m_\phi < 2.8$ GeV it should be lower than about 10\%. Motivated by this, in section~\ref{sec:future-LHC-direct}, we discuss the search for higher mass $\phi$ particles from Higgs decays as a promising future prospect at the LHC.

\subsection{Other constraints}

%%%%%%%%%%%%%%%%%%%%%%%%%%%%%%%%%%%%%%%%%%
\begin{figure}[t]
\begin{center}
\centering\includegraphics[angle=0, width=10.5cm]{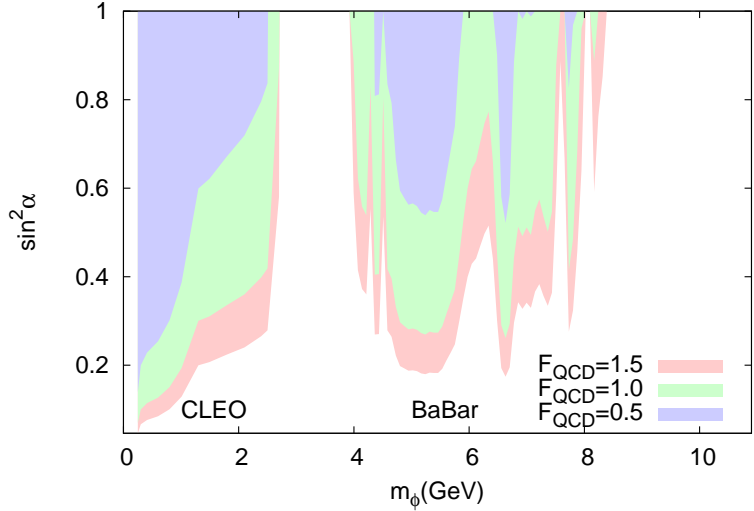}
\caption{\small \sl Bounds on the mixing angle $\sin^2 \alpha$ as a function of $m_\phi$, using constraints from $\Upsilon$(1S) decays in the BaBar and CLEO experiments. The correction factor $F_{QCD}$, as described in the text, has been varied in the range 0.5 to 1.5.}
\label{fig:upsilon}
\end{center}
\end{figure}
%%%%%%%%%%%%%%%%%%%%%%%%%%%%%%%%%%%%%%%%%

Light scalar particles can also be searched from the radiative decays of the bottom quark bound state $\Upsilon$. The decay of $\Upsilon \rightarrow \gamma \phi$ has been investigated for the mass region $m_{\phi} < m_{\Upsilon}$ in different experiments and the signal relies on a narrow peak of width around 10 MeV in the photon spectrum. At the leading order, the ratio of the partial widths for $\Upsilon \rightarrow \gamma \phi$ and $\Upsilon \rightarrow \mu^{+} \mu^{-}$ is given by 
\begin{equation}
R_0 =
\frac{\mbox{Br}(~\Upsilon \to \gamma \phi~)}
{\mbox{Br}(~\Upsilon \to \mu^{+} \mu^{-}~)} =
\frac{G_F m^2_b}{\sqrt{2} \pi \alpha_{em}}
\left( 1 - \frac{m^2_{\Phi}}{M^2_{\Upsilon} } \right) \sin^2 \alpha,
\label{eq:upsilon}
\end{equation}
where $G_F$ is the Fermi constant, $\alpha_{em}$ is the fine structure constant and $\alpha$ is the usual Higgs-singlet scalar mixing angle defined in the previous section. The numerator of equation~(\ref{eq:upsilon}), i.e., ${\rm Br}(~\Upsilon \rightarrow \gamma \phi~)$ receives multiple corrections, namely, QCD radiative correction, bound state correction and relativistic correction. The overall correction factor can be large enough~\cite{Gunion:1989we}. The correction term $F_{QCD}$ can be treated as a multiplicative factor in the right hand side of equation~(\ref{eq:upsilon}). Similarly, the denominator in equation~(\ref{eq:upsilon}) should be replaced by an expression including higher order corrections. However, the $\Upsilon$(1S) branching ratio to muons is an experimentally well measured quantity which we use for our calculations.

The most recent experimental limits on $\Upsilon (1S) \rightarrow \gamma \phi $ are obtained from the CLEO~\cite{Love:2008aa} and BaBar~\cite{Lees:2012te} in the $\phi \rightarrow \mu^{+} \mu^{-}$ and $\phi \rightarrow \tau^{+} \tau^{-}$ channels. No narrow peak in the photon spectrum has been observed by either experiment, except for the $\Upsilon \rightarrow J/\psi \gamma$ peak, which allows us to put bounds on scalar decay branching fractions to muons or taus. We take the 90\% C.L. upper limits from both experiments, and translate it to a limit on the ($\sin^{2}\alpha$, $m_{\phi}$)-plane as shown in figure~\ref{fig:upsilon}. We do not explicitly calculate the correction term $F_{QCD}$. Instead, we have varied it from 0.5 to 1.5 to estimate the sensitivity of our results on this factor. We find that in the very low mass range, the CLEO bound on $\phi \rightarrow \mu^{+} \mu^{-}$ is important, whereas in the high mass region ($m_{\phi} >$ 3 GeV ) the BaBar limit is similar to or slightly stronger than the CLEO limit. For this reason, we do not show the CLEO limit separately in the high mass region. Values of $\sin^2 \alpha$ below 0.1 is not constrained by $\Upsilon$(1S) decay even with a large positive correction factor ($F_{QCD}$=1.5).

We note in passing that, for very low values of the mediator mass, beam dump experiments can put constraints on the model parameters~\cite{Bjorken}. In such experiments, the scalar mediator $\phi$ can be emitted from electron or proton beams via bremsstrahlung, and if the mediator is very light, it may decay only after crossing the usually placed absorbing blocks. As the mixing angle of $\phi$ with the Higgs becomes smaller, although the mediator lifetime increases, but at the same time the production cross section goes down. In our scenario the beam dump constraints can be relevant in the very low mass region for $\phi$ ($\sim$ 0.1-0.5 GeV). For a detailed discussion on such constraints we refer the reader to reference~\cite{Zurek}.

%%%%%%%%%%%%%%%%%%%%%
%%%%% Prospects %%%%%
%%%%%%%%%%%%%%%%%%%%%
\section{Future prospects}
\label{sec: prospects}

%%%%%%%%%%%%%%%%%%%%%%%%%%%%%%%%%%%%%%%%%%
\begin{figure}[p]
\begin{center}
\centering\includegraphics[angle=0, width=14cm]{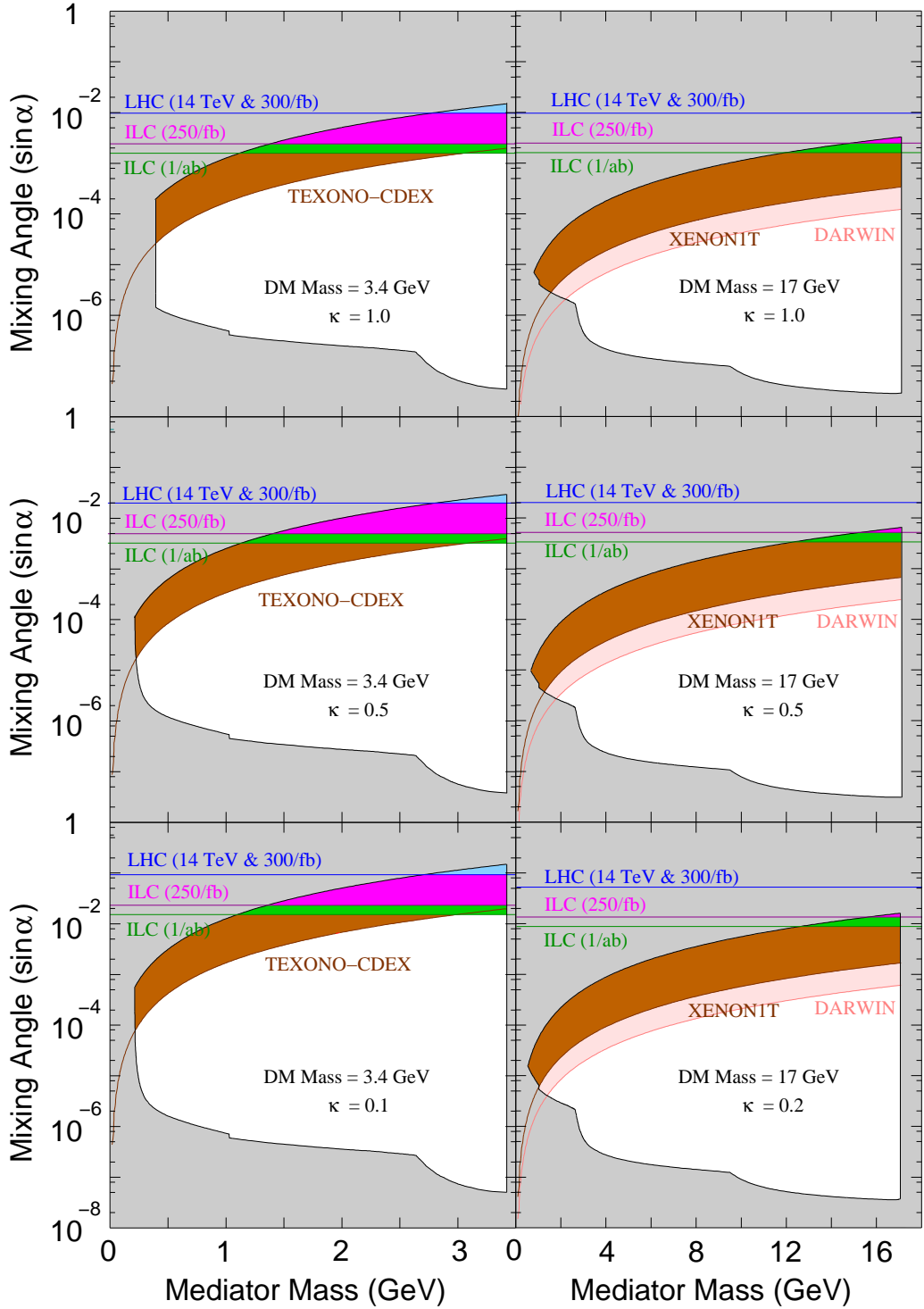}
\caption{\small \sl Expected reach in the ($m_\phi$, $\sin\alpha$) parameter space of the ADM model from LHC and ILC probes on the Higgs boson invisible branching ratio, as well as the reach of future dark matter direct detection experiments XENON1T, DARWIN and TEXONO-CDEX. The grey shaded area is excluded by current constraints (see figure \ref{fig: P}).}
\label{fig: F}
\end{center}
\end{figure}
%%%%%%%%%%%%%%%%%%%%%%%%%%%%%%%%%%%%%%%%%

We now estimate the prospects of future experiments to probe the ADM model parameter space which is allowed by current constraints described in the previous section.

\subsection{Future collider prospects}

\subsubsection{Higgs property measurements at the LHC (14 TeV) and ILC}

Estimates of the accuracy of Higgs couplings and invisible branching ratio measurement at the 14 TeV LHC with $300$ fb$^{-1}$ integrated luminosity have been performed in reference~\cite{Peskin}, according to which the $1\sigma$ upper limit on the invisible branching ratio is $5 \%$ (see also~\cite{Klute}). As discussed in further detail in reference~\cite{Peskin}, it is not clear whether these bounds can be significantly improved upon using the high-luminosity upgrade of the LHC, primarily because of increased systematic uncertainties. As a conservative estimate, a 10 \% upper bound at 95 \% C.L. on the invisible branching ratio would translate to an improvement of the previously obtained LHC bound in equation~(\ref{HiggsLHC}) as follows, \\
\underline{14 TeV LHC, 300 fb$^{-1}$ data:}
\begin{equation}
\sin^2\alpha <
\begin{cases}
\, 1.0 \, \kappa^{-2} \times 10^{-4} \qquad
\left( \, m_\chi = 17.1 \, {\rm GeV} \, \right),
\\
\, 9.4 \, \kappa^{-2} \times 10^{-5} \qquad
\left( \, m_\chi = 3.4 \, {\rm GeV} \, \right).
\end{cases}
\end{equation}

Precision measurement of the Higgs boson couplings to different SM final states is possible at the proposed International Linear Collider (ILC) experiment. Moreover, using the $e^+e^- \rightarrow Z h$ channel, a model-independent measurement of the Higgs branching ratio to invisible final states can be performed. Note that, in contrast to the global analysis of the LHC data, this measurement of the invisible branching ratio is independent of the Higgs boson production cross section.

At the 500 GeV ILC, with an integrated luminosity of 500 fb$^{-1}$, one can study the WW fusion production of Higgs, which, when combined with the individual branching ratios measured in the $Zh$ channel at the 250 GeV ILC, gives us an absolute determination of the total Higgs boson width, $\Gamma_{\rm tot}^h$. The accuracy expected is $\Delta \Gamma_{\rm tot}^h/\Gamma_{\rm tot}^h \simeq 6\%$~\cite{ILC-TDR}. Assuming that the measurement yields a central value which is equal to the SM width $\Gamma_{\rm tot}^{\rm SM}$ of 4.21 MeV, one can translate this into a bound for $\sin^2\alpha$ for a given value of $\kappa$ and $m_\chi$ as follows,\\
\underline{ILC $\Gamma_{\rm tot}^h$ measurement:}
\begin{equation}
\sin^2\alpha <
\begin{cases}
\, 5.7 \, \kappa^{-2} \times 10^{-5} \qquad
\left( \, m_\chi = 17.1 \, {\rm GeV} \, \right),
\\
\, 5.1 \, \kappa^{-2} \times 10^{-5} \qquad
\left( \, m_\chi = 3.4 \, {\rm GeV} \, \right).
\end{cases}
\end{equation}
~

On accumulating 250 fb$^{-1}$ data at the 250 GeV ILC, the projected accuracy (at 95\% C.L.) of the Higgs invisible branching ratio measurement using the Z-recoil spectra in $Zh$ production is 4.8\% using only the $Z \rightarrow \mu^+ \mu^-$ mode~\cite{ILC-TDR}. This is expected to improve significantly after including other visible decay modes of Z, including $Z \rightarrow q \overline{q}$, reaching a precision of $0.7 \%$~\cite{ECFA2013}. Assuming a $0.7 \%$ accuracy, we find that the ILC can probe the parameter space of our model down to very low values of the mixing angle,\\
\underline{ILC BR($h \to {\rm invisible}$) measurement:}
\begin{equation}
\sin^2\alpha<
\begin{cases}
\, 6.6 \, \kappa^{-2} \times 10^{-6} \qquad
\left( \, m_\chi = 17.1 \, {\rm GeV} \, \right),
\\
\, 5.9 \; \kappa^{-2} \times 10^{-6} \qquad
\left( \, m_\chi = 3.4 \, {\rm GeV} \, \right).
\end{cases}
\end{equation}
There is a possibility to improve upon this constraint further by measuring the invisible BR to $0.3 \%$ using $1$ ab$^{-1}$ data~\cite{private}, which would then lead to an upper bound on $\sin^2\alpha$ of $2.8 \times 10^{-6} \kappa^{-2}$ ($2.5 \times 10^{-6} \kappa^{-2}$) for $m_\chi=17.1$ ($3.4$) GeV. Thus, among all the future collider measurements, the ILC Higgs invisible BR determination seems to be the most effective one in constraining the mixing angle. 

In figure~\ref{fig: F}, we have shown the reach of the 14 TeV LHC and the ILC invisible Higgs branching ratio measurements in probing the ($\sin \alpha$, $m_\phi$) parameter space that is allowed after taking the current constraints into account. We find that these experiments can probe a part of the allowed region in the $m_\chi=3.4$ GeV case, and for higher values of $m_\phi$, they can compete with or even do better than the future direct detection probes like TEXONO-CDEX.

\subsubsection{Direct search of light scalar at 14 TeV LHC}
\label{sec:future-LHC-direct}

We have seen in section~\ref{sec:LHC-mediator} that the LHC limits on direct $\phi$ production are not very strong yet and the search channels rely only on the $\phi \rightarrow \mu \mu$ decay mode. We also saw that the bound in the $4 \mu$ channel, which is based on $\phi$ production from Higgs decays, is much stronger for lower values of $m_\phi$. We expect that the CMS and ATLAS collaborations will update their results in the near future in both these channels and will be able to probe the parameter space further. However, since the $\phi \rightarrow \mu^+ \mu^-$ branching ratio is very small above the charm quark threshold, $\phi$ masses above this threshold are not yet constrained by the LHC searches.

\begin{figure}[t]
\begin{center}
\centering\includegraphics[angle=0, width=10.5cm]{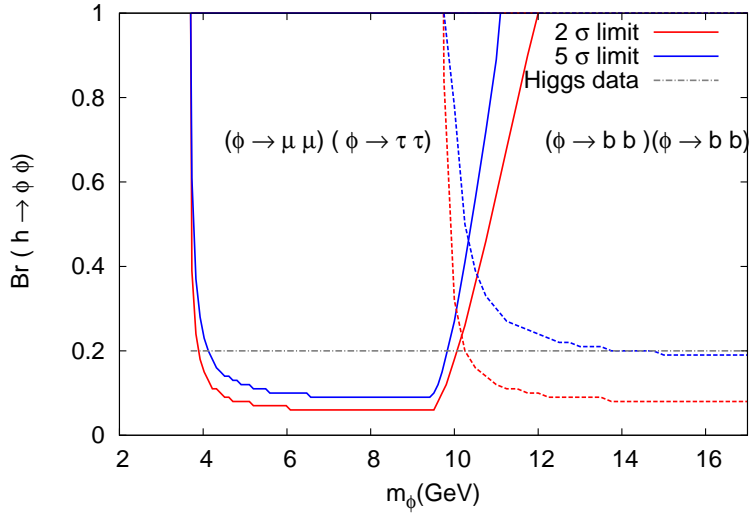}
\caption{\small \sl Expected future reach in the ($m_\phi$, BR($h \to \phi \phi$))-plane at the 14 TeV LHC with 300 fb$^{-1}$ integrated luminosity using the $b \bar{b}$, $\tau^+ \tau^-$ and $\mu^+ \mu^-$ decay modes of $\phi$. The solid and dashed lines correspond to the $2\mu 2\tau$ and $4b$ final states respectively. The horizontal dot-dashed line indicates the upper limit obtained from current LHC Higgs data.}
\label{fig:future}
\end{center}
\end{figure}

The other relevant channels to look for involve the $\phi \rightarrow \tau^+ \tau^-$ and $\phi \rightarrow b \bar{b}$ decay modes. In reference~\cite{Lisanti:2009uy}, the possibility to discover a light pseudo-scalar ($a$) particle in Higgs decays at the 14 TeV LHC has been studied, and the mass range relevant for this analysis is $2m_{\tau} < m_a <2 m_b$. The final state considered is 2$\tau$2$\mu$ with missing transverse energy coming from the boosted neutrinos in tau decays. The 95\% exclusion limit on $\sigma(g g \rightarrow h ) \times ~{\rm Br}(h \rightarrow a a )$ as a function of Higgs mass has been obtained for a pseudo-scalar mass of 7 GeV, assuming 2 ${\rm Br}(a \rightarrow \tau \tau) \times ~{\rm Br} (a \rightarrow \mu \mu) = 0.8 \%$. Using this result, we have estimated the reach on the ($m_\phi$, Br($h \rightarrow \phi \phi)$)-plane, which is shown in figure~\ref{fig:future}. We find that it is possible to probe $h \rightarrow \phi \phi$ branching ratios of up to about 5\% using 300 fb$^{-1}$ of data at the 14 TeV LHC.

%%%%%%%%%%%%%%%%%%%%%%%%%%%%%%%%%%%%%%%%%%
%%%%%%%%%%%%%%%%%%%%%%%%%%%%%%%%%%%%%%%%%

In reference~\cite{Carena:2007jk}, associated Higgs production with $W/Z$ at the 14 TeV LHC and its subsequent decay to light pseudo-scalars has been studied, in the context of the next to minimal supersymmetric standard model. The pseudo-scalar masses considered are above 10 GeV, and hence, it can dominantly decay to either b quarks or to $\tau$ leptons. After a detailed analysis, they show that the $4b$ final state is more promising than the $2b2\tau$ channel, and therefore, we concentrate only on the former channel. We have appropriately translated the expected reach as obtained in reference~\cite{Carena:2007jk}, and show in figure~\ref{fig:future} the future prospects to detect the scalar mediator in the ADM model using this mode. We find that for $m_\phi > 12$ GeV, it is possible to exclude $h \rightarrow \phi \phi $ branching ratios of up to 10\% using the 300 fb$^{-1}$ integrated luminosity at the 14 TeV LHC (a $5\sigma$ discovery of BR's allowed by the current LHC Higgs data seems to be difficult with this luminosity using the $4b$ mode). We should caution the reader that in translating the above bounds for our model, we have ignored possible effects of systematic uncertainties that might be present in the background estimates. 

For small values of the mixing angle, another interesting possibility is the production of a long-lived mediator at the LHC. Since the direct $\phi$ production rate will also be small in this case, the only way to probe such a scenario will be $\phi$ production from Higgs decays. We leave the detailed study of such a scenario for future work.

\subsection{Future reach of DM direct detection experiments}

In this sub-section, we briefly discuss the capability of future direct detection experiments to probe the allowed parameter space of the ADM model. We focus on three proposed experiments, namely, XENON1T~\cite{XENON1T}, DARWIN~\cite{DARWIN} and 
TEXONO-CDEX~\cite{TEXONO-CDEX}. XENON1T is the next generation liquid Xenon detector at the ton scale. This experiment aims to reduce backgrounds by a factor of 100 using established techniques and the sensitivity on the spin independent dark matter-nucleon scattering is expected to reach $\sim 5 \times 10^{-46}$ cm$^2$ for $m_{\chi} = 20$ GeV. Looking further into the future, a multi-ton noble liquid detector DARWIN is in the developmental stage. The expected mass of this liquid Xenon/Argon detector will be about 20 ton and it may reach a cross section range of $\sim 10^{-47}$ cm$^2$ for a 20 GeV dark matter. However, it is difficult to improve the sensitivity further, as solar neutrinos will start appearing as an irreducible background at these small values of cross section. Therefore, an efficient discrimination between electron-neutrino and dark matter-nucleon recoils will become crucial. In the case of a very light dark matter ($m_{\chi} =3.4$ GeV), we do not expect any improvements from either XENON1T or DARWIN, and detectors with a threshold below a keV will be required. A proposed future Germanium detector-based experiment, TEXONO-CDEX, expects to achieve a threshold of 100 eV, and is therefore ideal for a dark matter mass of a few GeV. The expected sensitivity in cross section for a 3 GeV dark matter is at least three orders of magnitude better than the TEXONO 2007 results. We show the expected reach from the XENON1T and DARWIN experiments in the $m_\chi=17.1$ GeV case, as well as that of TEXONO-CDEX for the $m_\chi=3.4$ GeV case in figure~\ref{fig: F}.

%%%%%%%%%%%%%%%%%%%
%%%%% Summary %%%%%
%%%%%%%%%%%%%%%%%%%
\section{Summary}
\label{sec: summary}

We have considered a light fermionic ADM scenario, where the mass of the ADM particle is determined by its charge under a gauged U(1)$_{\rm B-L}$. We discussed two example cases, with B$-$L charge of 1/3 and 5/3, in order to cover two interesting mass regions for the DM particle $\chi$. To obtain a sufficient annihilation cross section in the early universe, required to eliminate the symmetric component of the ADM, we considered the coupling of the ADM to a real singlet scalar field. This mediator scalar mixes with the SM Higgs boson after electroweak symmetry breaking. The DM-singlet coupling $\kappa$ is constrained by relic density considerations, and we obtained a lower bound of $\kappa > 0.1$ (0.2) for $m_\chi=3.4$ (17.1) GeV. The most stringent constraint in the mixing angle ($\sin \alpha$) and mediator mass ($m_\phi$) parameter space is obtained from DM direct detection experiments, XENON100 for the 17.1 GeV case, and TEXONO for the 3.4 GeV case. The observed properties of the Higgs boson at the LHC put an upper limit on its non-standard branching ratio, of the order of 20\%. This translates to an upper bound on $\sin^2\alpha$ which was found to be of $\mathcal{O}(10^{-4})$. This constraint is competitive with the TEXONO limit for higher values of $m_\phi$ and $m_\chi=3.4$ GeV. Furthermore, the CMS search for light scalars produced in Higgs decays in the $4 \mu$ channel, already limits the $h \rightarrow \phi \phi$ branching ratio to a level of 1\% for $m_\phi < 1.6$ GeV. Constraints coming from LEP $Z\phi$ search and precision electroweak observables, as well as from the measurement of $\Upsilon(1S)$ branching ratio are found to be rather weak compared to the above ones. We also derived a lower bound on the mixing angle ($\sin \alpha \gtrsim 10^{-6}$--$10^{-7}$ depending on $m_\phi$) from considerations of kinetic equilibrium between the DM sector and the SM sector in the early universe. 

After presenting the current allowed parameter space in the ADM model, we go on to estimate the expected reach from future experiments. The LHC 14 TeV run, with $300$ fb$^{-1}$ of data, will be able to probe $\sin^2\alpha \sim \mathcal{O}(10^{-5})$ using global fits of the data including a Higgs invisible branching ratio. The ILC can probe the Higgs invisible channel directly, with a reach of $\sin^2\alpha \sim \mathcal{O}(10^{-6})$. In addition, we also explored the possibilities of searching for the scalar mediator produced in Higgs decays at the 14 TeV LHC, especially for higher values of $m_\phi$, where the decays to bottom quarks or tau leptons dominate. We find that, depending upon the decay mode, Br($h \rightarrow \phi \phi$) of the order of 5-10\% can be probed using the $300$ fb$^{-1}$ data. The future DM direct detection experiments of XENON1T and DARWIN will have increased sensitivity for the higher mass DM case, while the proposed TEXONO-CDEX experiment can have a much better reach for the lower mass case. We have estimated the parameter region that can be covered by these experiments, which is in the range $\sin^2\alpha \sim \mathcal{O}(10^{-8}-10^{-10})$ . 

Even though we can probe down to very small mixing angles in future experiments, the lower bound on $\sin\alpha$, as dictated by the condition for kinetic equilibrium between the DM and SM sectors, is very difficult to explore. A {\em naturally} expected range for the mixing can be obtained if one speculates the existence of a $Z_2$ symmetry,  $\phi \rightarrow -\phi$, which is spontaneously broken by the VEV of the $\phi$ field. This will then induce a mixing term in the scalar mass matrix, which is fixed by dimensionless parameters in the scalar potential, and the VEV's of $\phi$ and the Higgs field. The mixing $\sin \alpha$ is then expected to be of $\mathcal{O}(10^{-2}-10^{-4})$, for $\mathcal{O}(1)$ values of the dimensionless parameters.

Throughout our study we have considered the case of a light CP-even scalar mediator. Instead, if one considers a pseudoscalar portal, the constraints from dark matter direct detection experiments become considerably weakened. On the other hand, the collider search strategies discussed in our study remain equally effective for a pseudoscalar. Therefore, irrespective of the CP properties of the scalar, the searches at colliders are going to be important,  while the direct detection probes will be crucial in covering a larger region of parameter space for the CP even case.

%%%%%%%%%%%%%%%%%%%%%%%%%%%
%%%%% Acknowledgments %%%%%
%%%%%%%%%%%%%%%%%%%%%%%%%%%
\section*{Acknowledgments}

This work is supported by the Grant-in-Aid for Scientific Research from the Ministry of Education, Science, Sports, and Culture (MEXT), Japan (Nos. 22244021 \& 23740169 for S.~Matsumoto, and No. 23104006 for M.M.~Nojiri), and also by the World Premier International Research Center Initiative (WPI Initiative), MEXT, Japan.

%%%%%%%%%%%%%%%%%%%%%%
%%%%% References %%%%%
%%%%%%%%%%%%%%%%%%%%%%

\end{document}